\documentclass[aps,prb,twocolumn,superscriptaddress,showpacs,preprintnumbers
]{revtex4-1}	
\usepackage[ngerman,english]{babel} 
\usepackage[utf8]{inputenc}

\usepackage{xcolor,soul}

\usepackage{amsmath,amssymb} 

\usepackage[pdfstartview=FitH,      
            breaklinks=true,        
            bookmarksopen=true,     
            bookmarksnumbered=true  
            ]{hyperref}				
            
\usepackage{wrapfig}
\usepackage{subfig}   

\usepackage{adjustbox} 		
\usepackage{upgreek}		
\usepackage{xargs}                      

\usepackage[colorinlistoftodos,prependcaption,textsize=tiny]{todonotes}
\newcommandx{\unsure}[2][1=]{\todo[linecolor=red,backgroundcolor=red!25,bordercolor=red,#1]{#2}}
\newcommandx{\change}[2][1=]{\todo[linecolor=blue,backgroundcolor=blue!25,bordercolor=blue,#1]{#2}}
\newcommandx{\info}[2][1=]{\todo[linecolor=OliveGreen,backgroundcolor=OliveGreen!25,bordercolor=OliveGreen,#1]{#2}}
\newcommandx{\improvement}[2][1=]{\todo[linecolor=Plum,backgroundcolor=Plum!25,bordercolor=Plum,#1]{#2}}
\newcommandx{\thiswillnotshow}[2][1=]{\todo[disable,#1]{#2}}

\begin{document}



\title{Low offset frequency $1/f$ flicker noise in spin torque vortex oscillators}


\author{Steffen Wittrock}\email[]{steffen.wittrock@u-psud.fr}\affiliation{Unit\'{e} Mixte de Physique CNRS/Thales, Univ. Paris-Sud, Univ. Paris-Saclay, 1 Avenue Augustin Fresnel, 91767 Palaiseau, France }
\author{Sumito Tsunegi}\affiliation{National Institute of Advanced Industrial Science and Technology (AIST),
Spintronics Research Center, Tsukuba, Ibaraki 305-8568, Japan}
\author{Kay Yakushiji}\affiliation{National Institute of Advanced Industrial Science and Technology (AIST),
Spintronics Research Center, Tsukuba, Ibaraki 305-8568, Japan}
\author{Akio Fukushima}\affiliation{National Institute of Advanced Industrial Science and Technology (AIST),
Spintronics Research Center, Tsukuba, Ibaraki 305-8568, Japan}
\author{Hitoshi Kubota}\affiliation{National Institute of Advanced Industrial Science and Technology (AIST),
Spintronics Research Center, Tsukuba, Ibaraki 305-8568, Japan}
 \author{Paolo Bortolotti}\affiliation{Unit\'{e} Mixte de Physique CNRS/Thales, Univ. Paris-Sud, Univ. Paris-Saclay, 1 Avenue Augustin Fresnel, 91767 Palaiseau, France }
 \author{Ursula Ebels}\affiliation{ Univ. Grenoble Alpes, CEA, INAC-SPINTEC, CNRS, SPINTEC, 38000 Grenoble, France }
 \author{Shinji Yuasa}\affiliation{National Institute of Advanced Industrial Science and Technology (AIST),
Spintronics Research Center, Tsukuba, Ibaraki 305-8568, Japan}
\author{Gilles Cibiel}\affiliation{Centre National d'\'{E}tudes Spatiales (CNES), 18 av. Edouard Belin, 31401 Toulouse, France}
\author{Serge Galliou}\affiliation{FEMTO-ST Institute, CNRS, Univ. Bourgogne Franche Comt\'{e}, 25030 Besançon, France}
\author{Enrico Rubiola}\affiliation{FEMTO-ST Institute, CNRS, Univ. Bourgogne Franche Comt\'{e}, 25030 Besançon, France}
 \author{Vincent Cros}\email[]{vincent.cros@cnrs-thales.fr}\affiliation{Unit\'{e} Mixte de Physique CNRS/Thales, Univ. Paris-Sud, Univ. Paris-Saclay, 1 Avenue Augustin Fresnel, 91767 Palaiseau, France }


\date{May 13, 2019}		
\preprint{ { \begin{minipage}{\columnwidth} {\raggedleft PHYSICAL REVIEW B \textbf{99}, 235135 (2019) }   \end{minipage} } \hfill {\begin{minipage}{\columnwidth} \begin{flushright} DOI: \href{https://doi.org/10.1103/PhysRevB.99.235135 }{10.1103/PhysRevB.99.235135}  \end{flushright} \end{minipage}}   }

\begin{abstract}

Low frequency noise close to the carrier remains little explored in spin torque nano oscillators. However, it is crucial to investigate as it limits the oscillator's frequency stability.
This work addresses the low offset frequency flicker noise of a TMR-based spin-torque vortex oscillator in the regime of large amplitude steady oscillations.
We first phenomenologically expand the nonlinear auto-oscillator theory aiming to reveal the properties of this noise. 
We then present a thorough experimental study of the oscillator’s $1/f$ flicker noise and discuss the results based on the theoretical predictions. Hereby, we connect the oscillator’s nonlinear dynamics with the concept of flicker noise and furthermore refer to the influence of a standard $1/f$ noise description based on the Hooge formula, taking into account the non-constant magnetic oscillation volume, which contributes to the magnetoresistance.

\end{abstract}

\pacs{}

\maketitle

\section{Introduction}

Spin torque nano oscillators (STNOs) convert magnetization dynamics into electrical rf signals by exploiting fundamental phenomena of spintronics, such as the magnetoresistive effect \cite{Fert1988,Gruenberg1989} and the concept of spin transfer torque \cite{Berger1996,Slonczewski1996}, which allows sustaining their dynamics.
Since 2003, the investigation of STNO properties \cite{Kiselev2003} has attracted a huge interest as they have been considered not only as a unique opportunity for studying nonlinear dynamics at the nanoscale, but also as promising candidates for next-generation multifunctional spintronic devices \cite{Locatelli2013}. 
More recently, STNOs have also been demonstrated to be capable of being key elements in the realization of broadband microwave energy harvesting \cite{Fang2018} or frequency detection \cite{Jenkins2016,Louis2017}, and of reconstructing bio-inspired networks for neuromorphic computing \cite{Torrejon2017,Romera2018}. \\
These possible applications are all strongly linked to the STNO's fundamental property, i.e. its nonlinearity, {which describes a coupling between the oscillator's amplitude and phase{\cite{Kim2008,Slavin2009}}.}
It provides the basis to  manifold phenomena such as injection locking to an external rf signal \cite{Lebrun2015,Hamadeh2014}, synchronization of multiple STNOs \cite{Kaka2005,Locatelli2015,Lebrun2017} or the spin torque diode effect \cite{Jenkins2016,Tulapurkar2005,Miwa2013,Naganuma2015}. 
However, nonlinear behaviour also causes the oscillator's poor spectral coherence and leads to a conversion from amplitude to phase noise \cite{Kim2008,Slavin2009,Grimaldi2017}, which limits its amplitude and frequency stability, and therefore its applicability in real practical devices.  

Noise mechanisms in tunnel- or giant magnetoresistive effect (TMR/GMR resp.) based magnetic sensors, which represent the building block of a STNO,  are usually the frequency independent shot (only for TMR) and thermal noise, and low frequency $1/f$-noise, also called flicker noise. In certain cases, Random Telegraph Noise (RTN), arising from the fluctuations between two different energy levels, can also exist. 
Its sources are mainly of electronic or magnetic origins and equally present in STNOs as well, where it is anticipated that the existence of self sustained nonlinear magnetization dynamics drastically influences the noise characteristics compared to classical magnetic sensors.\\
The STNO's noise distribution for offset frequencies far from the carrier frequency is now reasonably well understood. At room temperature, it has been found that it is mainly attributed to dominant thermal noise in the framework of nonlinearity \cite{Grimaldi2017,Quinsat2010,Slavin2009,Silva2010}. However, the noise at low offset frequencies, which is crucial for most of the targeted applications, remains largely under investigation.
The existence of flicker noise at these low frequency offsets is of a general nature, as noise of spectral $1/f^{\beta}$-scaling is present in a wide variety of physical systems, such as stellar emissions, lake turbulences, Nile flooding, and virtually all electronic devices \cite{Rubiola2008}. 
Initially recognized by Johnson \cite{Johnson1925}, and later investigated in thin films \cite{Bernamont1937}, it has recently attracted attention as the dominant noise source in GMR and TMR sensors \cite{Nowak1999,Arakawa2012,Almeida2006,Feng2012} and their major drawback in terms of performance.\\
In STNOs, the $1/f$ flicker noise at low offset frequencies has so far only been experimentally recognized\cite{Keller2010, Eklund2014,Quinsat2010,Grimaldi2017}. {In Refs. {\cite{Quinsat2010,Grimaldi2017}}, the focus has been on thermal noise at high offset frequencies and  an indication of colored $1/f$ noise at lower offsets has been found. Ref. {\cite{Keller2010}} presents a measurement method to access the $1/f$ regime and shows its influence on the frequency spectra. In Ref. {\cite{Eklund2014}}, the study concerns $1/f$ noise in the very specific geometry of nano-contact STNOs. A large influence on multiple present modes, which are particularly avoided within the work presented here, has been found.   }
{In the present work, we systematically investigate the $1/f$ flicker noise in vortex based STNOs (in the following called STVOs) both experimentally, and theoretically based on the corresponding nonlinear Langevin equations {\cite{Slavin2009}} together with a phenomenological approach. Moreover, we present a thorough study of the important parameters governing this type of noise.  }

In magnetic sensor applications, the flicker noise, which can not yet be described in its entire universality, is typically described by the empirical Hooge formalism \cite{Hooge1969,Fermon2013}. In this work, we also aim at establishing a connection between a magnetic TMR sensor, a fundamental building block of the STNO, and the latter's intrinsic nonlinearity.

We focus our study on vortex based spin torque nano-oscillators (STVOs). 
The fundamental mode in these devices, the gyrotropic vortex motion \cite{Cowburn1999,Shinjo2000}, is deterministically well understood and can be described by reducing the complex dynamics only to the dynamics of the vortex core (the so-called Thiele formalism \cite{Thiele1973}). Moreover, it is energetically well separated from other modes reducing mode crossing probabilities. 
STVOs, compared to other STNOs exploiting different magnetic modes (as e.g. a uniform precession or local modes), exhibit large amplitude oscillations with frequencies from $100\,$MHz up to $\sim 1.5\,$GHz, a rather narrow linewidth of $\sim 100\,$kHz, and output powers of up to a few $\upmu$W \cite{Tsunegi2014}. 
Since magnetic vortex dynamics has in many aspects been considered as a model system for the study of magnetic dynamics \cite{Grimaldi2017}, STVOs are suitable to fundamentally study the noise behaviour of STNOs. 
{We believe that the general results of this work can be extrapolated to every kind of STNO, as the nonlinear nature and the basic noise generating processes are the same. Thus, the main objective is to establish a connection between flicker noise and the STNO's nonlinearity and to study the STNO's flicker noise in general. Such study has not been performed yet for STNOs, even if this is an important issue for the different types of applications aiming to rely on them.

The article is organized in the following way: We first present the basic mechanisms of the fundamental gyrotropic mode in spin torque vortex oscillators (STVOs). We connect the latter with the general nonlinear auto-oscillator theory, which provides a theoretical basement for the description of spin torque nano oscillators (STNOs) in general. From this model, we deduce the STNO's noise behaviour especially including low offset frequency $1/f$ flicker noise processes via a phenomenological approach. In section {\ref{sec:experiment}}, we describe the experiment's details and give basic properties of our samples. In section {\ref{sec:results_noise_STVOs}}, we present experimental data showing the noise characteristics in STVOs and subsequently, in section {\ref{sec:flicker-noise_in_STVOs}}, focus on the low offset frequency regime. We compare and analyze the experimental data regarding the developed model and, in section {\ref{sec:flicker-noise_in_STVOs}}, introduce the formalism of \textsc{Hooge} to discuss the governing parameters of flicker noise in STNOs in general and STVOs in particular.  }

\section{General Framework of magnetization dynamics and noise in STNOs and STVOs}
\label{sec:general_framework}

\subsection{Noise properties of the STVO's gyrotropic mode}

 In laterally confined geometries such as disks, an excitation of the vortex core from its equilibrium position at the center of the circular nano pillar leads to a radial motion in the sub-GHz regime around the equilibrium, i.e. the gyrotropic mode of the vortex core. Assuming the vortex core as a soliton and subsequently applying Thiele's approach \cite{Thiele1973}, this gyrotropic mode is mainly characterized by the vortex core position $\vec{X}(t)$, which is   
 represented by the oscillation's orbit radius $r(t)$, i.e. its normalized value $s(t)=r(t)/R$,  and its phase $\phi (t)$.
 Fig. \ref{fig:Forces} schematically summarizes the forces, which mainly define the gyrotropic motion within the Thiele formalism \cite{Thiele1973,Dussaux2012}: The gyroforce, arising from the vortex non collinear profile, 
 the damping force expressing the magnetic relaxation, the confinement force due to the magnetostatic energy, and the spin transfer force.

As the injected spin transfer torque overcomes the damping, the vortex core autooscillates on an isoenergetic trajectory (fig. \ref{fig:vortex-motion}). Due to the system's nonlinear nature, it is always pushed onto a stable limit cycle as the oscillation amplitude $a (t) \sim s (t)$ becomes lower than the stable value and is pulled back as it becomes higher.

\definecolor{hellblau}{RGB}{48,157,181}
\definecolor{orange}{RGB}{228,108,10}
\begin{figure}[bth!]
  \centering  
  \subfloat[
  \label{fig:Forces}]
  {  
  \includegraphics[width=0.39\columnwidth]{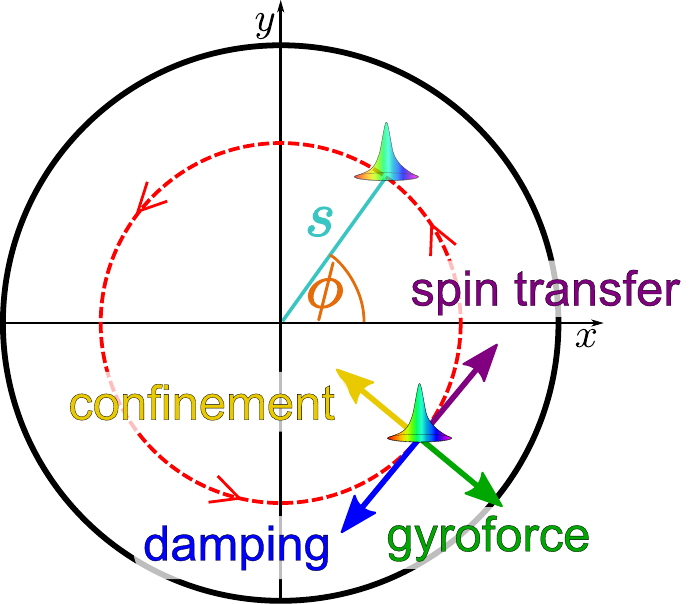}  }   
  \hfill 
 \subfloat[
  \label{fig:noise-types}] 
  { 
  \includegraphics[width=0.44\columnwidth]{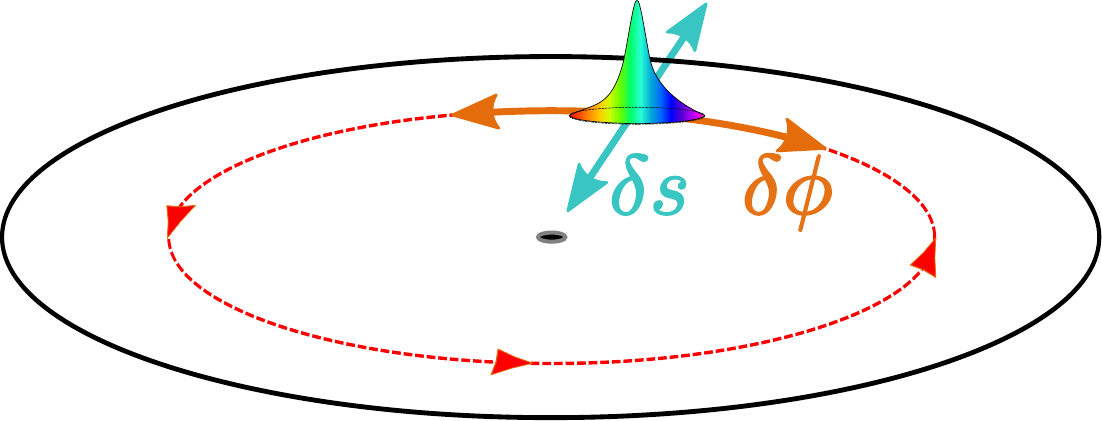}  } 
  
  \caption[]{(a) Schematics of forces acting on the magnetic vortex core: gyroforce (green arrow), confinement force (yellow), spin transfer (purple), and damping force (blue); (b) Noise schematics of vortex motion:  amplitude and phase noise \textcolor{hellblau}{$\delta s$} and \textcolor{orange}{$\delta \phi$} resp.} 
  \label{fig:vortex-motion}
\end{figure}

This limit cycle mechanism also appears in presence of noise, which perturbs the ideal trajectory and is separated into amplitude and phase noise, as depicted in fig. \ref{fig:noise-types}.
Noise sources can be manifold and have recently been mainly attributed to thermal origins \cite{Kim2008,Tiberkevich2007,Kudo2009,Georges2009,Quinsat2010,Grimaldi2017}, which furthermore lead to a change in spectral shape and an increase in linewidth of the oscillation \cite{Grimaldi2017}. Moreover, the presence of colored $1/f$ noise has been shown experimentally  at offset frequencies close to the carrier \cite{Keller2010, Eklund2014,Quinsat2010,Grimaldi2017}.

\subsection{The general nonlinear auto-oscillator theory}
\label{subsec:Slavin-theory}

We first remind the general nonlinear auto-oscillator theory\cite{Slavin2009,Tiberkevich2007,Slavin2008}, which is a universal theoretical approach for nonlinear oscillators in general and spin torque nano oscillators (STNOs) in particular.
For STNOs, this model can be derived from the Landau-Lifshitz-Gilbert-Slonczewski equation and describes the deterministic dynamics of the nonlinear oscillator through the complex oscillation amplitude $c(t)=\sqrt{p(t)}e^{i\phi (t)}$:
\begin{align}
\frac{dc}{dt} + i\omega(|c|^2)c + \Gamma_+ (|c|^2)c - \Gamma_- (|c|^2)c = f(t) ~~~. \label{eq:Langevin-eq}
\end{align}
{In this model, the gyrotropic motion of the vortex core (as described above) can be identified through oscillation amplitude $a(t) = \sqrt{p(t)}$ and phase $\phi(t)$ of the complex variable $c(t)$.}
$\omega$ hereby denotes the oscillation's angular frequency, $\Gamma_+$ the positive damping rate representing the losses of the system, and $\Gamma_-$ the negative damping rate representing the system's gain. $f(t)$ is a phenomenological term, which allows a description of the system's interaction with the environment. In the case regarded here, that mainly includes different noise-processes, among them thermal and flicker noise, which make equation (\ref{eq:Langevin-eq}) a nonlinear stochastic Langevin equation. \\
Due to the dependence of the damping parameters on the amplitude ($\frac{d\Gamma_-(p)}{d p} < 0$ and $\frac{d\Gamma_+(p)}{d p} > 0$), the oscillation is described by a limit cycle with stable oscillation power $p_0= |c|^2$, which is obtained when the positive and negative damping terms equal: $\Gamma_-(p)=\Gamma_+(p)$. 
Assuming perturbation $\delta p$ of the stable oscillation power due to noise gives a characteristic damping rate $\Gamma_p = \pi f_p = \left[ \frac{d\Gamma_+}{dp} (p_0) - \frac{d\Gamma_-}{dp} (p_0) \right] p_0$ of small power deviations back to the stable trajectory. 
The parameter $\nu = N p_0/(\pi f_p)$ with $N=d\omega(p)/dp$ the nonlinear frequency shift coefficient is the normalized dimensionless nonlinear frequency shift and quantifies the coupling between phase and amplitude due to nonlinearity.

\subsection{Theoretical description of low offset frequency flicker noise}

In the following, we assume a phenomenological ansatz to describe the low offset frequency noise in the framework of the nonlinear autooscillator theory. Its parameters can for STVOs be identified with the deterministic exploitation of the Thiele equation on the spin transfer torque induced gyrotropic motion\cite{Dussaux2012,Grimaldi2017}. {More detailed steps of the following derivation can be found in the supplementary information\cite{supplementary_noise_PRB_2018}.}

We define $f_n(t)$, which is the realization of the stochastic $1/f$ noise process, to act independently on amplitude and phase: $g_n(t) = \Re\left[ f_n(t) e^{-i\phi (t)} \right] $ and $h_n(t) = \Im\left[ f_n(t) e^{-i\phi (t)} \right] $. Here, the flicker noise process is assumed to be stationary and so are $g_n$ and $h_n$. Note that the colored $1/f$ noise can be modeled from an ensemble of statistically independent Ornstein-Uhlenbeck processes with varying correlation times\cite{Kaertner1990}.
Linearization of eq. (\ref{eq:Langevin-eq}) with the perturbed oscillator power $p=p_0 + \delta p$  and the phase fluctuations $\delta \phi (t)$   gives the coupled differential equations \cite{Slavin2009,Kim2008,Silva2010}:
\begin{align*}
\frac{d (\delta p)}{dt} + 2\pi f_p \delta p(t) = 2 \sqrt{p_0}~ g_n (t) \\
\frac{d(\delta \phi)}{dt} = \frac{1}{\sqrt{p_0}} ~ h_n(t) - N ~ \delta p (t) ~~~.
\end{align*}
Solving in the frequency space yields:
\begin{align*}
\delta p(f) &= \frac{\sqrt{p_0}}{\pi f_p + i\pi f} \cdot g_n (f) 
\intertext{and}
\delta \phi (f) &= \frac{1}{2 \pi i f \sqrt{p_0} } ~ h_n(f) - \frac{N}{2\pi i f} ~ \delta p (f) ~~~.
\end{align*}

Inserting the experimental $1/f^{\gamma}$-behaviour of the noise spectral density into the autocorrelation's Fourier transform exploiting the Wiener-Khintchine-theorem
\begin{align*}
S_{f_n}(f) = \int \langle f_n(t)f_n^*(t-t')\rangle \cdot e^{2\pi i f t} dt = \frac{\alpha}{f^{\gamma}} ~~~,
\end{align*}
\hspace{0.25cm} allows formulating the expression for the PSDs of phase and amplitude noise in the low offset frequency regime.
\definecolor{hellblau}{RGB}{48,157,181}
\definecolor{orange}{RGB}{228,108,10}
Together with the noise PSD resulting from thermal noise processes (as studied in Ref. \cite{Grimaldi2017,Quinsat2010,Slavin2009,Kim2008,Silva2010}), we obtain the total noise PSD expression ${S_{\delta \epsilon}}$ for the normalized amplitude $\epsilon = a/a_0 = s/s_0$ and  ${S_{\delta \phi} } $ for the phase respectively:
\begin{align}
{S_{\delta \epsilon}} &= \frac{\Delta f_0}{\pi} \cdot \frac{1}{f^2 + f_p^2} + \frac{1}{4 p_0 \pi^2 \left( f_p^2 + f^2   \right)} \cdot  \frac{\alpha}{f^{\gamma}} \label{eq:all-noise_amp} \\
{S_{\delta \phi} } &=  \frac{\Delta f_0}{\pi f^2} + \frac{1}{4 p_0 \pi^2 f^2 } \cdot \frac{\alpha}{f^{\gamma}} + \frac{\nu^2 f_p^2}{f^2} {S_{\delta \epsilon}}  ~~~.   \label{eq:all-noise_phase}
\end{align}
The first terms in both equations, proportional to the oscillation linewidth $\Delta f_0$, correspond to thermal origins, whereas all other terms describe the additional PSD due to flicker noise, which is at the center of the present study.
Note that the variable $\alpha$ for the amplitude noise is not necessarily equivalent to $\alpha$ in the phase noise equation as different mechanisms might contribute.

The flicker noise terms describe the original noise of exponent $\gamma$, which leads to a $1/f^{\gamma}$ slope for the amplitude and a $1/f^{\gamma+2}$ slope for the phase noise.
{Equation ({\ref{eq:all-noise_phase}}) clearly shows the nonlinear conversion from amplitude to phase noise.
 Moreover, we also find an additional}, non-coupled pure phase flicker noise term -- the second term in eq. ({\ref{eq:all-noise_phase}}) --, which also exhibits a $1/f^3$-characteristic.


\section{Experiment}
\label{sec:experiment}

The samples are circular shaped magnetic tunnel junctions {of diameter $2R=375\,$nm}, which consist of a pinned layer made of a conventional synthetic antiferromagnetic stack (SAF), a MgO tunnel barrier and a FeB free layer {exhibiting a magnetic vortex configuration}. The layers are deposited by ultrahigh vacuum (UHV) magnetron sputtering and patterned using e-beam lithography and Ar ion etching. After annealing at $360\,^{\circ}$C for $1\,$h, the resistance-area product is $RA\approx 4\,\Omega\cdot\upmu m^2$.
The TMR ratios lie around $100\,$\% at room temperature {with resistance values of $1/G_0=(44-45)\,\Omega$ in the vortex state, slightly dependent on the applied current with negative slope}.  
The SAF is composed of \selectlanguage{ngerman} PtMn($15$)/""Co$_{71}$Fe$_{29}$($2.5$)/""Ru($0.86$)/""CoFeB($1.6$)/""Co$_{70}$Fe$_{30}$($0.8$) and the total layer stack is SAF/""MgO($1$)/""FeB($4$)/""MgO($1$)/""Ta/Ru, with the nanometer layer thickness in brackets.
\selectlanguage{english}
The SAF's top magnetic layer exhibits a uniform magnetization in the film plane, which can be slightly tilted by applying an external magnetic field $H_{\perp}$ perpendicular to the latter. {For all the presented measurements, $H_{\perp}$ is chosen to $\mu_0 H_{\perp}=500\,$mT}. This gives an out-of-plane spin polarization of the injected dc current providing the necessary spin transfer torque component to sustain the gyrotropic vortex motion \cite{Dussaux2010}. 
Increasing the dc current injected through the layer structure leads to a stronger spin transfer torque counteracting the natural damping of the gyrotropic vortex core motion until, at a certain threshold current value $I_{c}$, the damping is overcome and auto-oscillation of the vortex core around the disk-center arises \cite{Dussaux2010}.

The TMR effect allows the magnetization dynamics to be converted into an electrical rf signal.
In the circuit, the dc and rf current parts are separated through a bias tee and the measured rf signal is evaluated as the direct interpretation of the magnetization dynamics. 
 
\begin{figure}[bth!]
  \centering  
  \captionsetup[subfigure]{labelformat=empty}
  \subfloat{  
  \hspace{0.5cm} \includegraphics[width=0.68\columnwidth]{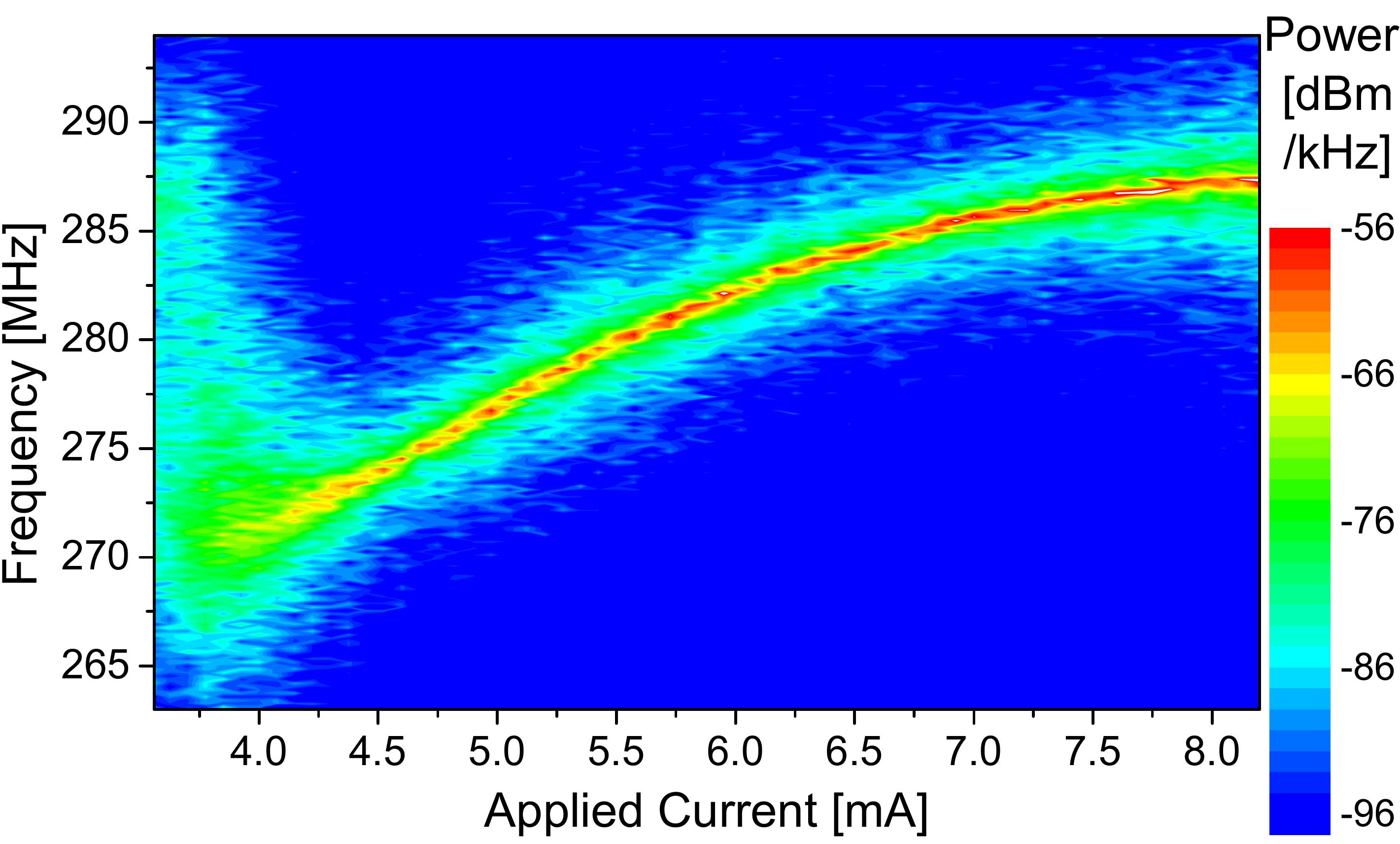}  }   
  \newline 
 \subfloat[\label{fig_hysterese:demag_unten}] 
  { 
  \includegraphics[width=0.9\columnwidth]{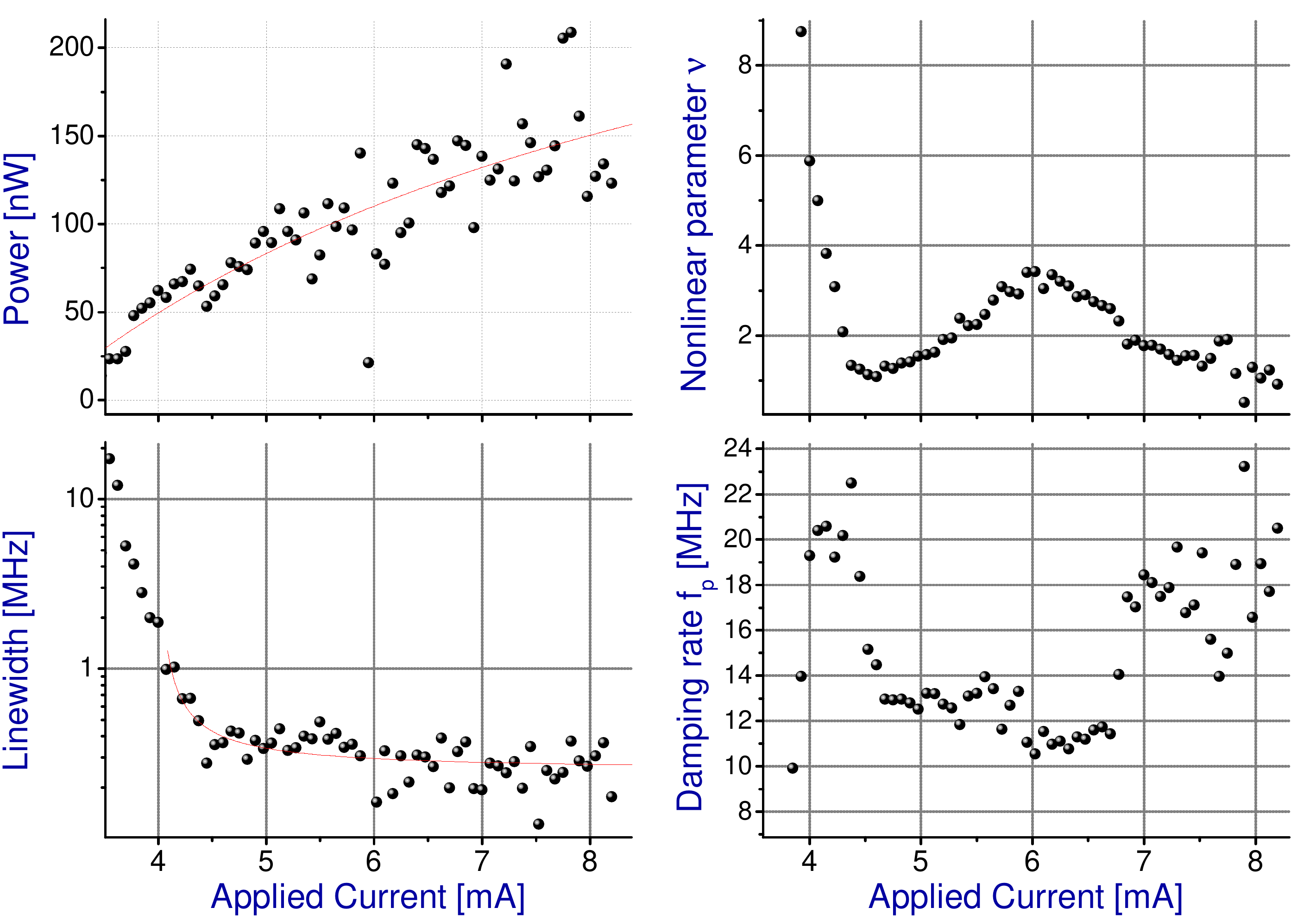}  } 
  
  \caption[]{Characteristic oscillation parameters, such as the oscillation frequency $f_c$, output power $p_0$, nonlinear parameter $\nu$, linewidth $\Delta f_0$, and damping rate $f_p$,  as a function of the applied current $I_{dc}$ at $\mu_0 H_{\perp}=500\,$mT. }
  \label{fig:basic-properties}
\end{figure}

The magnitude of the oscillation's output power $p_0$, its linewidth $\Delta f_0$, and the carrier frequency $f_c$ {(the frequency of the spin torque driven magnetization dynamics)} are obtained from Lorentzian fits of the emission power spectra{, measured by a spectrum analyzer}. 
Noise data are gathered from single-shot oscilloscope voltage time traces and evaluation via the Hilbert transform method\cite{Bianchini2010,Quinsat2010}. 
To obtain noise data close to the carrier (down to $1\,$Hz offset), the signal is first down-converted to $(2\pm 1)\,$MHz via frequency mixing with an external source (see Ref. \cite{Keller2010,Eklund2014}).{ This allows to decrease the oscilloscope's sampling rate (from $\sim 5\,$GSa/s down to $\sim 40\,$MSa/s) and to increase the measurement time (from $\sim 8.2\,$ms to $\sim 2.05\,$s)  while conserving the signal's noise properties.} High frequency components are filtered by a low pass filter {(bandwidth DC to $22\,$MHz).
More details on the noise measurement and data processing are given in the supplementary information\cite{supplementary_noise_PRB_2018}.} 
The oscillator's nonlinearity parameters (more explicitly described in section \ref{subsec:Slavin-theory}) are determined from the oscilloscope's time space measurements. The damping rate $f_p$ back to the stable oscillation trajectory is estimated by performing an exponential fit to the signal amplitude's autocorrelation\cite{Bianchini2010}. The nonlinearity parameter $\nu$, a measure for the system's nonlinearity, is obtained by comparing the coupling between amplitude and phase noise as elaborated in section \ref{sec:general_framework}. In fig. \ref{fig:basic-properties} we present the evaluated basic parameters of the STVO's oscillation at the chosen experimental conditions. 
 In addition, we fit the data with the expected behaviour of the magnitudes. The evolution of the power and linewidth with the applied current can be fitted from their theoretical description \cite{Slavin2009} $p_0(I)=(I/I_{c}-1)/(I/I_{c}+Q)\cdot p_{max}$ and $\Delta f_0(I)\sim 1/p_0(I)$ respectively, with the oscillation threshold current $I_{c}\approx 3.7\,$mA, and $Q$ and $p_{max}$ the respective fitting parameters. The parameter $Q$ represents the nonlinearity of the damping (cp. Ref. \cite{Slavin2009}).

In this work, we restrict the measurements and analysis to the oscillation behaviour in the dc current regime corresponding to the gyrotropic mode. In particular, we avoid
any perturbing effects of mode crossing (as in Ref. \cite{Eklund2014}) or more complex dynamical behaviour in the regime of very large excitation. Indeed beyond the stable gyrotropic mode (mode crossing, energy transfer into other subordinate modes), under specific current and field conditions, the occurrence of Random Telegraph Noise (RTN) at characteristic frequencies of up to $10\,$kHz can be observed. However, this is not further investigated here {(and even particularly avoided)}, as the detailed analysis of the noise characteristics in the RTN regime goes beyond the scope of this paper.


\section{Results - Noise in STVOs}
\label{sec:results_noise_STVOs}

A standard visualisation of noise properties is the representation of the power spectral density (PSD), which has the advantage to allow a direct distinction of the underlying noise processes and to classify them by their characteristic inverse power law behaviour $PSD \sim 1/f^{\beta}$.
In fig. \ref{fig:noise-curve_with_flicker-coupling}, {we show (among other curves explained later) typical measurement data for the phase noise (black curve) and the amplitude noise PSD (grey curve)} of the studied STVO sample {at an applied current of $I_{dc}=4.6\,$mA}. The noise PSDs are depicted versus the offset $f_{\text{off}}=|f-f_c|$ from the carrier frequency $f_c$ {(frequency of the vortex gyrotropic motion)}. 
The behaviour is mainly described by the nonlinear nature of the oscillator. For higher offset frequencies ($f_{\text{off}}\gtrsim 10^5\,$Hz in fig. \ref{fig:noise-curve_with_flicker-coupling}), the curve's shape is governed by thermal white gaussian noise processes, thoroughly investigated in Ref. \cite{Quinsat2010} for uniform STNOs (STNOs exploiting the uniform precession of magnetization dynamics) and in Ref. \cite{Grimaldi2017} for STVOs (STNOs exploiting the vortex gyrotropic mode).
In this region, {for clarity theoretically represented in fig. {\ref{fig:noise-curve_with_flicker-coupling}}'s inset based on eqs. ({\ref{eq:all-noise_amp}}-{\ref{eq:all-noise_phase}})}, the noise signature behaves linearly below the relaxation rate $f_p$ with exponent $\beta=2$ for the phase and $\beta=0$ for the amplitude noise. The coupling between the amplitude and phase leads to a conversion of amplitude to phase noise and therefore to an increase of the phase noise by $10 \log(1+\nu^2)$. 
For $f_{\text{off}}\gg f_p$, perturbations are faster than the nonlinear damping response $\tau_p = 1/(2\pi f_p)$, which is usually in the order of a few tens of oscillation periods (depending on the spin torque strength), so that the nonlinearity becomes less significant. The amplitude noise decreases with $1/f^2$ and the phase noise with $1/f^4$ for $f_p \ll f \ll \nu f_p$ and with $1/f^2$ for $f\gg\nu f_p$. At even higher offset frequencies the Johnson-Nyquist noise floor eventually limits the PSDs (not shown in fig. \ref{fig:noise-curve_with_flicker-coupling}).	

\definecolor{light-gray}{rgb}{0.96,0.96,0.96}

For low offset frequencies $f_{\text{off}} \lesssim 10^4\,$Hz, that are in the focus of this work, the PSD spectra exhibit a $1/f^3$- and a $1/f^1$-behaviour for phase and amplitude noise respectively, where $1/f$-flicker-noise is found to be dominant over thermal noise processes. The fits $\sim \alpha_{exp}/f^{\beta}$ on the experimental data in fig. \ref{fig:noise-curve_with_flicker-coupling} (orange curve for phase noise and cyan curve for amplitude noise) illustrate the curves' typical slopes and furthermore give the corresponding experimental noise prefactor $\alpha_{exp}$, later discussed in section \ref{sec:flicker-noise_in_STVOs}.\\
 The fundamental origin of flicker noise{, described through the phenomenological parameter $\alpha$ in eqs. ({\ref{eq:all-noise_amp}}-{\ref{eq:all-noise_phase}}),} cannot implicitly be specified as different physical processes are potential origins. This includes intrinsic phenomena such as fluctuations in the magnetic texture\cite{Diao2011,Nowak1999,Arakawa2012}, 
 the impact of defects and/or inhomogeneities in the magnetic layers or the tunnel barrier in particular due to the fabrication process\cite{Eklund2014}. Even external fluctuations of the driving dc current or the applied magnetic field can play a role. 

\begin{figure}[hbt!]
		\centering
		\vspace{-0em}
		\includegraphics[width=1.02\columnwidth]{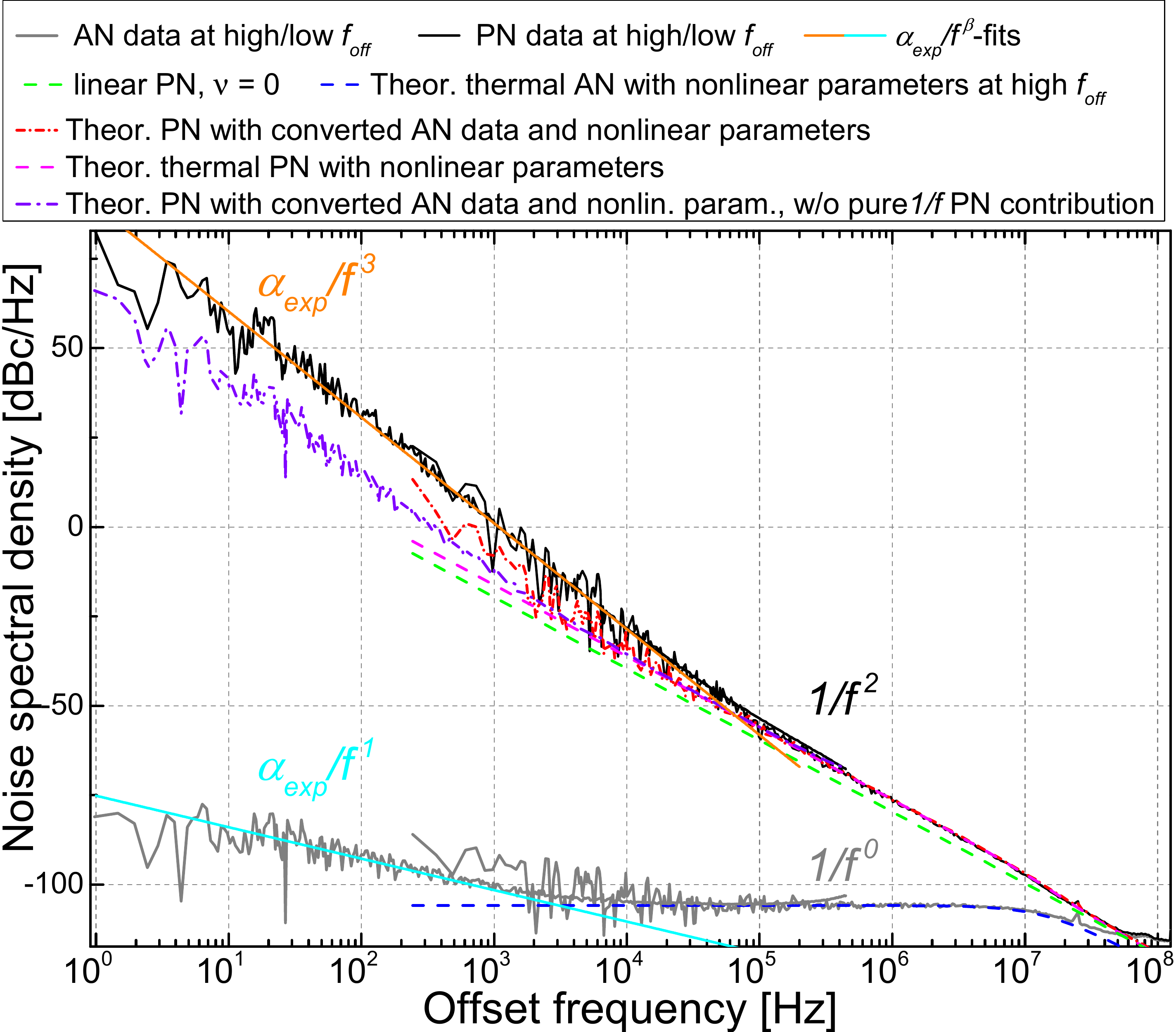}\llap{\raisebox{1.126\height}{\adjustbox{lap={\width}{0.398cm}}{{\setlength{\fboxsep}{0.3pt}\fcolorbox{black}{light-gray}{\includegraphics[width=0.452\columnwidth  ]{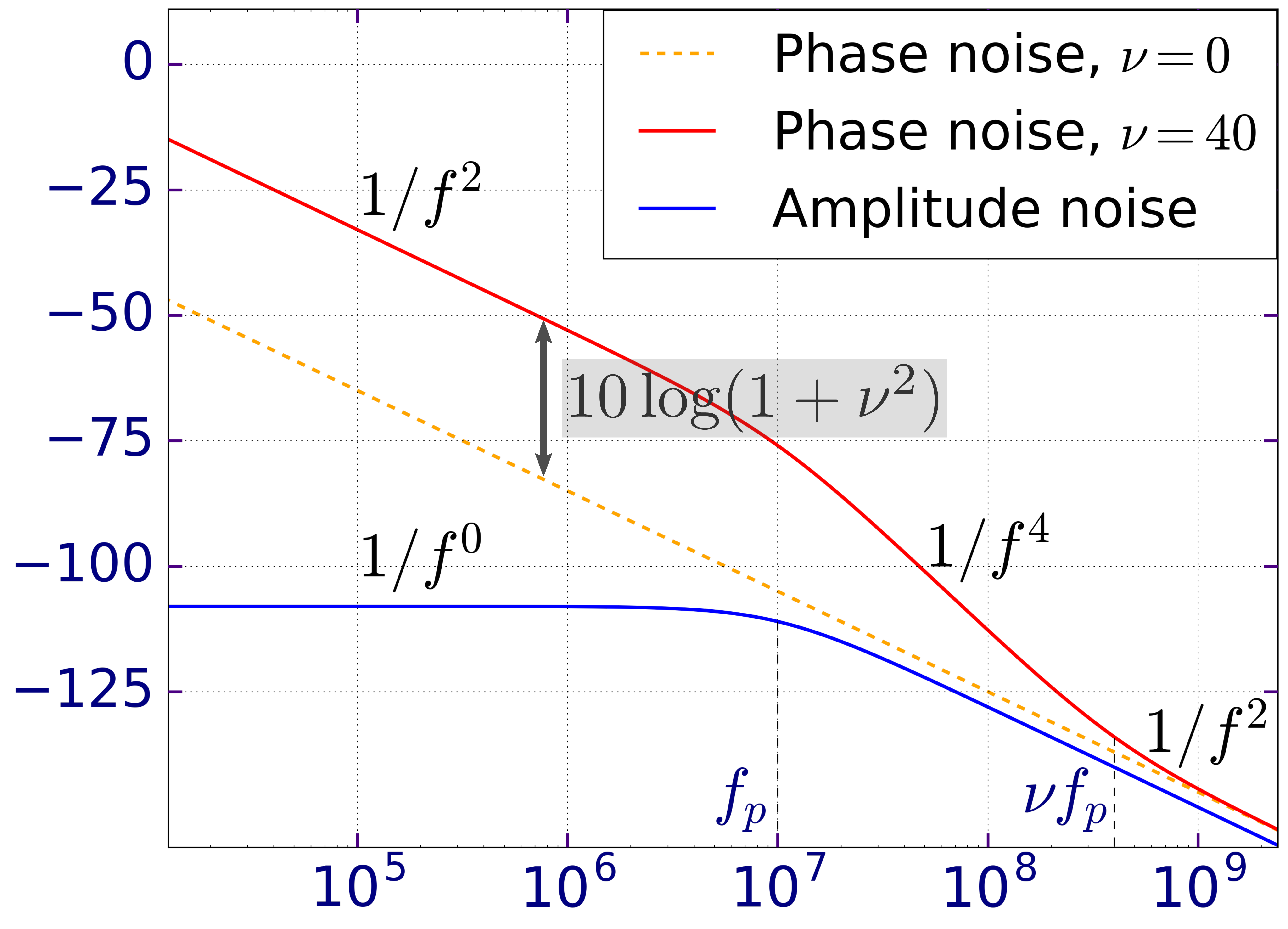} } } } }   } 
		\caption[]{PSD of phase (PN) and amplitude (AN) noise; $\mu_0 H_{\perp}=500\,$mT, $I=4.6\,$mA, as well as theoretical curves based on eqs. (\ref{eq:all-noise_amp}-\ref{eq:all-noise_phase}) and described in the text. The inset shows an analytical representation (eqs. (\ref{eq:all-noise_amp}-\ref{eq:all-noise_phase}), but different parameters as for the measurement) of the higher offset frequency noise due to thermal effects. The phase noise of a linear oscillator ($\nu=0$, dashed line) is shown as well as its increase due to nonlinearity.} 
		\label{fig:noise-curve_with_flicker-coupling}
		\end{figure}

Comparing the experimental data depicted in fig. \ref{fig:noise-curve_with_flicker-coupling} with the theoretical equations (\ref{eq:all-noise_amp}-\ref{eq:all-noise_phase}), we find for the generating noise process $\gamma \approx 1$ and a conversion of $1/f^1$ amplitude into $1/f^3$ phase noise, reflected both in fig. \ref{fig:noise-curve_with_flicker-coupling} (black and grey curve in the low offset frequency regime) as well as in eq. (\ref{eq:all-noise_phase}).
Next to the experimental PSDs, fig. \ref{fig:noise-curve_with_flicker-coupling} also displays {theoretical curves excluding one or more terms in equations} (\ref{eq:all-noise_amp}) and (\ref{eq:all-noise_phase}), and described in the following. The preliminarily evaluated experimental values $p_0$, $\nu$, $\Delta f_0$ and $f_p$, presented in fig. \ref{fig:basic-properties}, are fed into the equations. Only considering the thermal linear part of the phase noise PSD ($\nu = 0$, green curve in fig. \ref{fig:noise-curve_with_flicker-coupling}) again emphasizes the noise growth due to nonlinearity $\nu>0$, as this curve is clearly below the experimental one (black in fig. \ref{fig:noise-curve_with_flicker-coupling}).
For higher offset frequencies (flicker contribution negligible), theoretical amplitude and phase noise curves (blue dashed and pink dashed respectively) exhibit, as expected, an excellent agreement to the corresponding data (black and grey curves resp.).\\
For lower offset frequencies, from $\sim 10^4\,$Hz on downwards, the so far neglected flicker noise sums to the thermal noise parts and lets the experimental and theoretical curve diverge. The red dot-dashed curve in fig. \ref{fig:noise-curve_with_flicker-coupling} describes the theoretical phase noise PSD including the extra contribution from the conversion of experimental amplitude noise through the coupling term in eq. (\ref{eq:all-noise_phase}) -- thus it only neglects the pure flicker phase noise term (second term in eq. (\ref{eq:all-noise_phase})). Importantly, we note that this curve still clearly remains below the experimental data (black curve in fig. \ref{fig:noise-curve_with_flicker-coupling}) in the range of low offset frequencies, but fits well for higher offsets.

\section{Results - Analysis of Flicker Noise regime } 
\label{sec:flicker-noise_in_STVOs}

In fig. \ref{fig:noise-curve_with_flicker-coupling}, we demonstrate that the above described theoretical curves, based on the thermal noise contribution, correspond well to the experimental data at higher frequencies.
At lower offset frequencies ($f_{\text{off}} \lesssim 10^4\,$Hz), they clearly differ. 
To understand this difference, we further evaluate eq. (\ref{eq:all-noise_phase}) { using the low offset amplitude noise to additionally consider amplitude-phase coupling} and to classify the important parameters in the region $f_{\text{off}} \lesssim 10^4\,$Hz.
Hence, the violet curve in fig. \ref{fig:noise-curve_with_flicker-coupling} represents the phase noise only including the thermal and the amplitude-converted contribution (similar to the dot-dashed red curve in the high offset frequency regime). The difference between the experimental phase noise PSD (black curve in fig. \ref{fig:noise-curve_with_flicker-coupling}) and the calculated curve (violet) therefore clearly pinpoints the pure phase flicker noise in fig. \ref{fig:noise-curve_with_flicker-coupling}, which is found to be dominant against the other contributions. 
To further classify the important noise parameters, we concentrate on the pure phase flicker noise in the following.\\
\begin{figure}[hbt!]
		\centering
		\vspace{0em}
		\includegraphics[width=0.87\columnwidth]{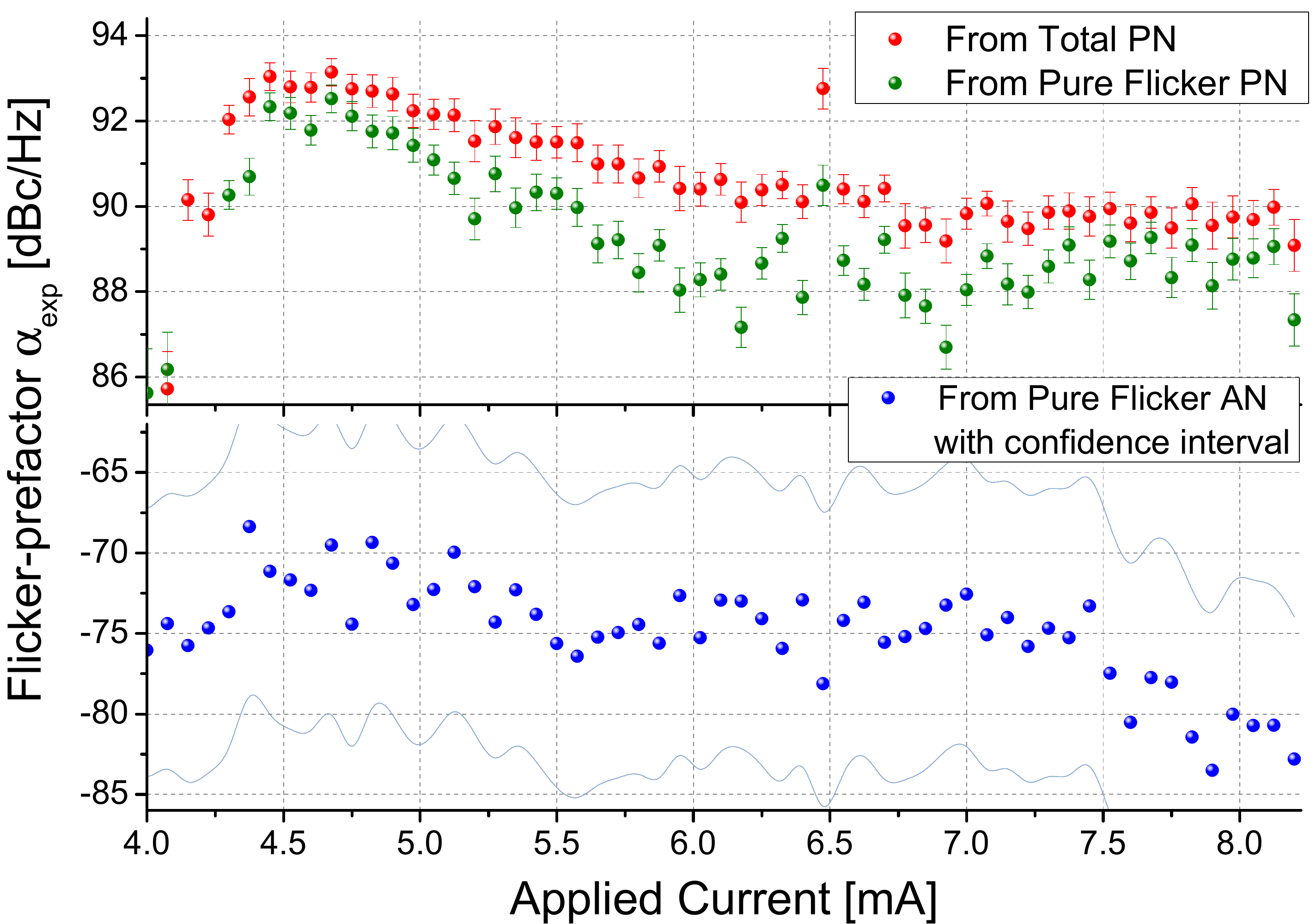} 
		\caption[]{Evolution with applied current of the prefactors $\alpha_{exp}$ from the $\alpha_{exp}/f^{\beta}$-fits of (i) the experimental phase noise (PN) data (red points) (ii) of the pure flicker PN (difference between black and violet curve in fig. \ref{fig:noise-curve_with_flicker-coupling}), represented by green points, and (iii) of the amplitude noise (AN) data (blue points). 
		}
		\label{fig:prefactors_vs_current}
		\end{figure}
In fig. \ref{fig:prefactors_vs_current}, we compare the experimental prefactors $\alpha_{exp}$ of the $\alpha_{exp}/f^{\beta}$-fits on the experimental low frequency noise as depicted in fig. \ref{fig:noise-curve_with_flicker-coupling} and plot them as a function of the injected dc current. 
The green points in fig. \ref{fig:prefactors_vs_current} represent the difference between the experimental data curve (black in fig. \ref{fig:noise-curve_with_flicker-coupling}) and the data curve including thermal noise and nonlinear amplitude noise conversion (violet in fig. \ref{fig:noise-curve_with_flicker-coupling}), notably the pure phase flicker noise. 
The difference between the two curves (red and green in fig. \ref{fig:prefactors_vs_current})
 directly reflects the contribution from the part of the noise which is converted from amplitude to phase.
Indeed, this difference is small and therefore emphasizes that the pure flicker PN is dominant for the flicker noise regime. This conclusion is one of the important results of the work.

\subsection*{Governing parameters and relation to the Hooge formalism}

Another important result is that the flicker noise for both the amplitude (blue points in bottom of fig. \ref{fig:prefactors_vs_current}) and the phase noise (red and green points in fig. \ref{fig:prefactors_vs_current}) is decreasing as the injected dc current is increased.
For comparison, this behaviour is different from what is found in conventional TMR sensors \cite{Nowak1999,Herranz2010,Fermon2013}, whose layer structure is very similar to the one of STNOs. For these devices, the low frequency noise behaviour is usually evaluated only by a description based on the phenomenological 
\textsc{Hooge}-formula \cite{Fermon2013,Hooge1969}:
\begin{align}
S_P = \frac{\alpha_H P_{dc}}{A\cdot f^{\beta}} ~~~, \label{eq:Hooge-proportionality}
\end{align}
\hspace{0.25cm} with $P_{dc}$ the circuit's dc power, $A$ the magnetically active surface, that is usually constant in TMR sensors, and $\alpha_H$ the Hooge-parameter. Interestingly, the coefficient $\alpha_H$ is often considered as a kind of quality factor e.g. in magnetic sensor technologies. {It includes the intrinsic origins of the flicker noise due to manifold potential mechanisms already mentioned before.\\
 Note that the phenomenological Hooge formula is originally well adapted to semiconductor devices {\cite{VanDerZiel1988,Sangwan2013,Balandin1999}}. There, it describes the $1/f$ noise proportional to the power supplied into the system averaged over the number of carriers. 
 By applying the Hooge model to magnetic sensors, the averaging is indeed realized on the effective magnetic volume that is converted into the active magnetic surface $A$ after proper renormalization in eq. ({\ref{eq:Hooge-proportionality}}).
In STVOs, the nonlinear evolution of the active surface $A$ or, more generally, the active magnetic volume has to be especially taken into account}, as usually not the entire device area is active, but only the surface enclosed by the vortex core trajectory contributes to the rf signal (red circle in fig. \ref{fig:vortex-motion}). A particularity of STVOs (and STNOs in general) is that, 
resulting from the nonlinearities of the different forces acting on the vortex core, $A$ can be controlled through the amplitude of the spin transfer torque. \\
The active surface $A=\pi (R s_0)^2$ can be experimentally determined by measuring the conductivities in the vortex or antiparallel state $G_0(I_{dc})$ and $G_{ap}(I_{dc})$, the TMR value $TMR(I_{dc})$, the corresponding applied field values $h_i$  normalized to the magnetization of the layer $i$, and the effective oscillation voltage $V_{rf}$ \cite{Grimaldi2017}. In the vortex oscillator model system, the oscillation radius can be calculated through:
\begin{align*}
s_0 =& \frac{2G_0 (1+ G_0 Z_{50\Omega})}{I_{dc}\cdot TMR(I_{dc}) \cdot G_{ap} Z_{50\Omega} \lambda} \\ & \cdot \frac{1}{\sqrt{ (1-h_{free}^2)(1-h_{SAF}^2)}} \cdot V_{rf0}   ~~~,
\end{align*}
\hspace{0.25cm} with $Z_{50\Omega}$ the load, and $\lambda \approx 2/3$ a magnetoresistive factor (proportionality factor for the planar magnetization value relative to the saturation magnetization). Note that the average resistance $1/G_0 =R_0(I_{dc})$ usually decreases for larger currents.

\begin{figure}[hbt!]
		\centering				
		\vspace{0em}				
		\includegraphics[width=0.86\columnwidth]{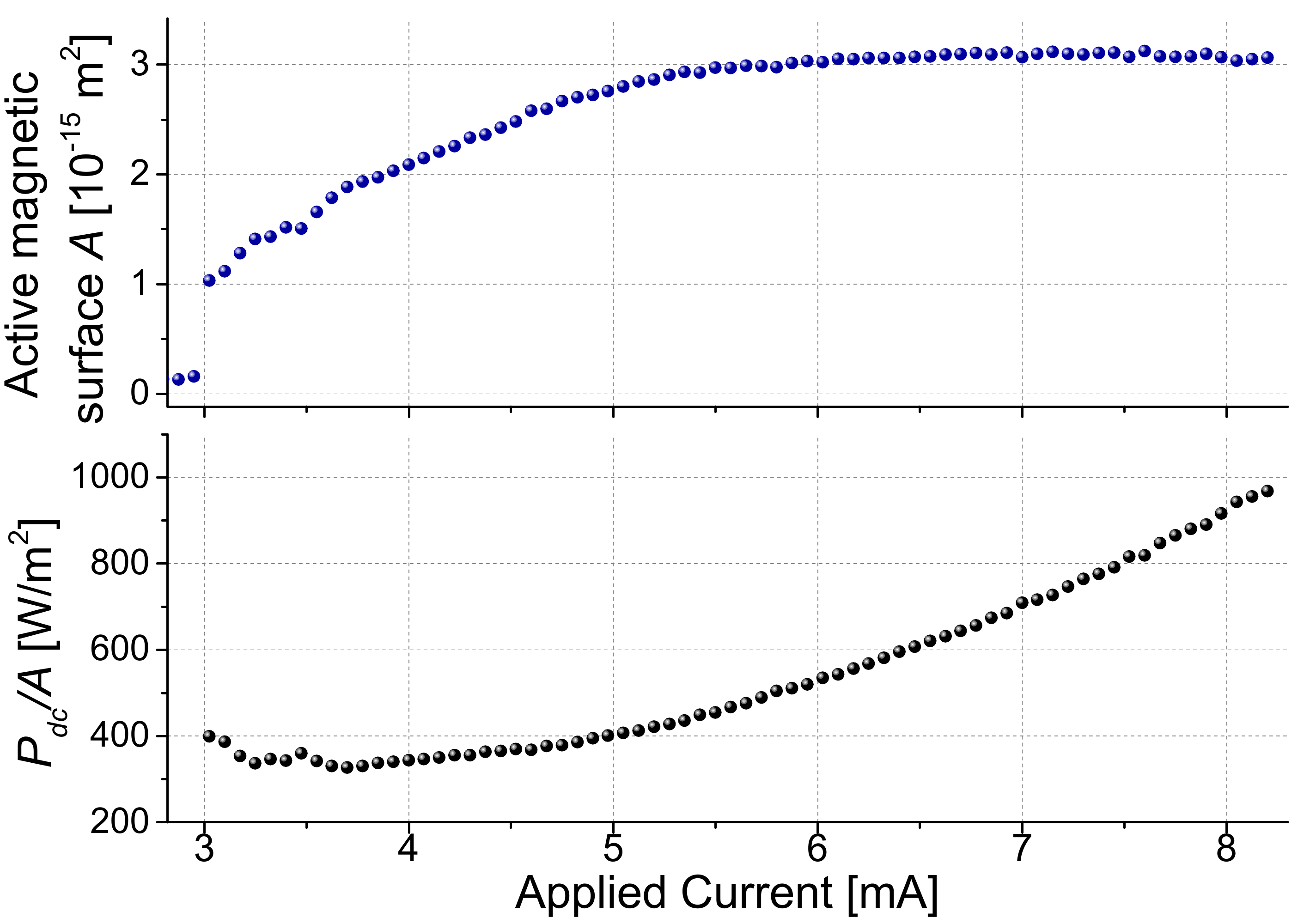}\llap{\raisebox{13.2\height}{\adjustbox{lap={\width}{-5.4cm}}{\fcolorbox{black}{light-gray}{ a) }  } }   }\llap{\raisebox{6.52\height}{\adjustbox{lap={\width}{-5.4cm}}{\fcolorbox{black}{light-gray}{ b) }  } }   }
		\caption[]{Experimental active magnetic surface $A$ (a) and  Hooge proportionality $P_{dc}/A$ (b) vs. applied dc current $I_{dc}$. }
		\label{fig:Volume_and_Hooge}
		\end{figure}

The evaluated active surface of the studied STVO as a function of the dc current is shown in fig. \ref{fig:Volume_and_Hooge}, as well as the Hooge-proportionality $P_{dc}/A$, which mainly determines the noise prefactor in eq. (\ref{eq:Hooge-proportionality}).

We find that the Hooge proportionality $P_{dc}/A$ is first constant and then increases as a function of $I_{dc}$. For lower currents, the evolution of $A$ is in fact compensating the increase. But then in the regime of constant $A$, the quadratic behaviour of $P_{dc}/A$ is similar to what is usually found in magnetic TMR/GMR sensors.
Following the Hooge model, a consequence is that also the noise level for higher currents shall increase. 
As shown in fig. \ref{fig:prefactors_vs_current}, this is contrary to what we find experimentally. Thus, we conclude, that $P_{dc}/A$ alone does not properly describe the flicker noise behaviour of a STVO, as it mainly does in case of magnetic sensors.

Indeed, the nonlinear nature of the oscillator, expressed by the equations (\ref{eq:all-noise_amp}-\ref{eq:all-noise_phase}), has to be taken into account.  It obviously dominates over the Hooge-behaviour, which is in fact valid for the static case with no sustained dynamics. 
In eq. (\ref{eq:all-noise_amp}-\ref{eq:all-noise_phase}) {we assume the generating noise (which is not deterministically understood so far) to follow Hooge's description in eq. ({\ref{eq:Hooge-proportionality}}) and include it in the parameter $\alpha \sim \alpha_H P_{dc}/A$.}  
Note that the prefactor $\alpha_H$ for the phase does not necessarily need to be equivalent to the one of the amplitude noise.

\begin{figure}[hbt!]
		\centering
		\vspace{0em}
		\includegraphics[width=0.86\columnwidth]{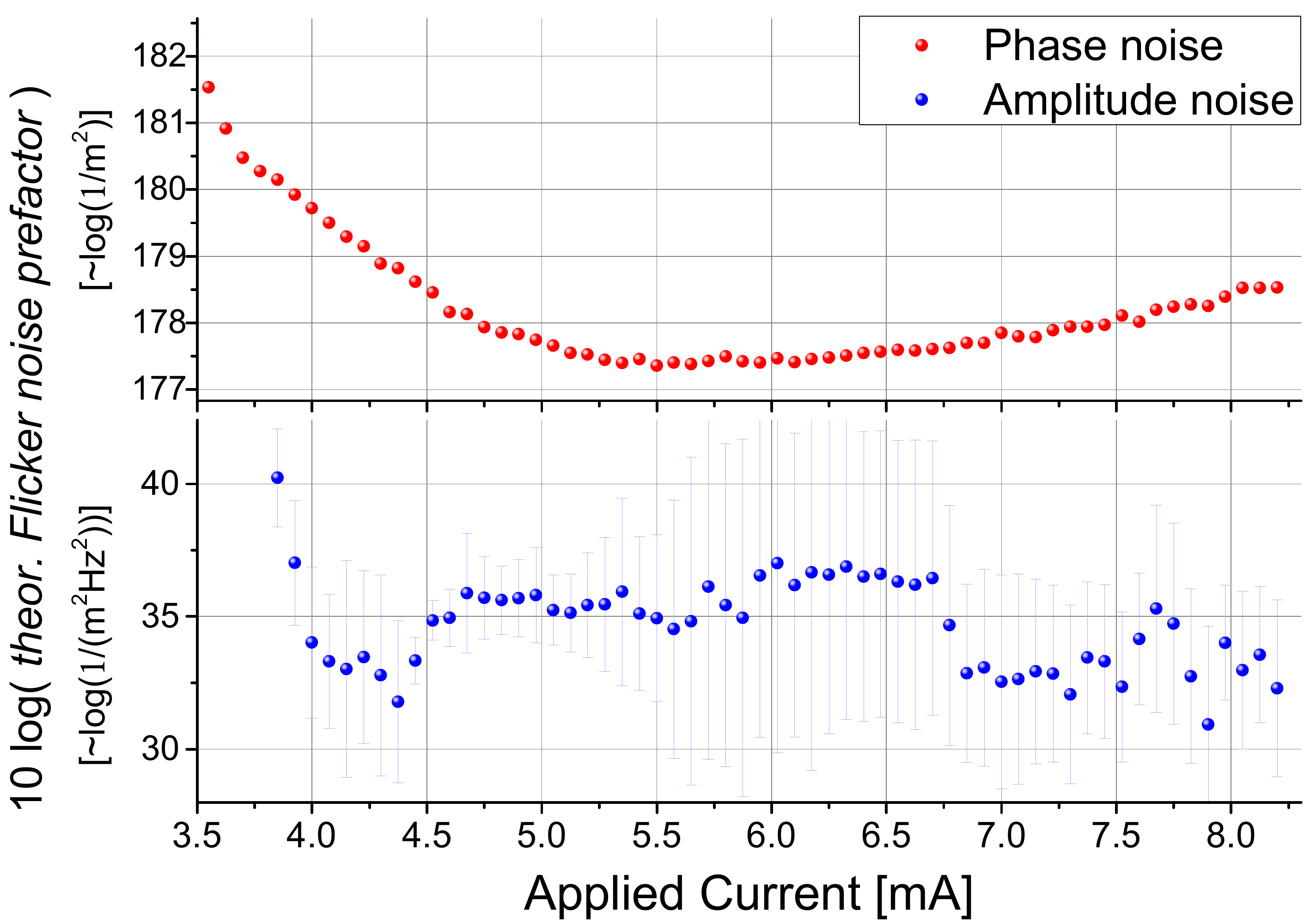} 
		\caption[]{Calculated pure flicker noise prefactors evaluated through eq. (\ref{eq:all-noise_amp}-\ref{eq:all-noise_phase}) with measured experimental magnitudes without $\alpha_H$. }
		\label{fig_theo_prefactors:prefactors}
		\end{figure}

In fig. \ref{fig_theo_prefactors:prefactors}, we calculate the evolution of the flicker noise prefactor vs. $I_{dc}$ based on the parameters of eq. (\ref{eq:all-noise_amp}-\ref{eq:all-noise_phase}) taking into account the previously discussed effects. At this stage, we neglect $\alpha_H$ which will be discussed later, and therefore plot $P_{dc}/(4p_0\pi^2 f_p^2 A)$ for the amplitude and $P_{dc}/(4p_0\pi^2 A)$ for the pure phase flicker noise prefactor.  The calculated noise prefactors of the pure amplitude and phase flicker noise terms (fig. \ref{fig_theo_prefactors:prefactors}) 
 qualitatively reproduce the trends of the direct measurement, represented in fig. \ref{fig:prefactors_vs_current}. 
In particular, we see that the phase noise curve first decreases, then remains constant over a certain current range until is starts increasing again from $\sim 6.5\,$mA on. This behaviour is associated with the decreasing nonlinearity at higher currents \cite{Grimaldi2017} and a resulting nearly constant oscillation radius $s_0$ and output power $p_0$ in this regime (cp. figs. \ref{fig:Volume_and_Hooge} and \ref{fig:basic-properties}). Therefore, the Hooge-ratio $P_{dc}/A$ also becomes more influential and starts to dominate the noise-behaviour similarly to what is usually found in TMR sensors.\\
The amplitude noise prefactor (fig. \ref{fig_theo_prefactors:prefactors}, blue curve) is slightly decreasing in the range of the experimentally studied injected current, but it has to be remarked that the error bars are large. 
In general, the flicker contribution of $S_{\delta \epsilon}$ mainly depends on the damping rate $f_p$ back to the limit cycle.

The Hooge parameter $\alpha_H$ (not yet considered for the data shown in fig. \ref{fig_theo_prefactors:prefactors}) can be estimated from our results as a proportionality factor between the calculated noise prefactors (shown in fig. \ref{fig_theo_prefactors:prefactors}) and the measured values (shown in fig. \ref{fig:prefactors_vs_current}). 
We estimate $\alpha_{H,\delta \phi} \approx 10^{3}\,\Omega\,\upmu$m$^2 \Leftrightarrow G_0 \alpha_{H,\delta \phi} \approx 10^1\,\upmu$m$^2$ for the phase and $\alpha_{H,\delta \epsilon} \approx 10^{1}\,\Omega\,\upmu$m$^2\Leftrightarrow G_0 \alpha_{H,\delta \epsilon} \approx 10^{-1}\,\upmu$m$^2$ for the amplitude noise. 
For comparison, in TMR based sensors the obtained values are typically between $G_0 \alpha_H\approx 10^{-6}\,\upmu$m$^2$ (Ref. \cite{Nowak1999,Gokce2006}) and $10^{-11}\,\upmu$m$^2$ (Ref. \cite{Arakawa2012,Almeida2008,Aliev2007,Scola2007}), measured for voltage fluctuations upon the tunnel junction. 
In Ref. \cite{Arakawa2012}, $G_0\alpha_H \approx 10^{-11}\,\upmu$m$^2$ is calculated from data on TMR sensor devices with a layer structure similar to ours and resistance-area product $RA$ well below $100\,$k$\Omega\cdot \upmu$m$^2$, where $\alpha_H$ usually becomes stable in terms of $RA$\cite{Fermon2013}. 
In contrast to the present work, data in Ref. \cite{Arakawa2012} were obtained at low temperature ($T=5\,$K). Thermal noise processes can also be the generator of $1/f$ noise processes (Barkhausen noise), as shown in Ref. \cite{Scola2007}, where a decrease of $\alpha_H$ by about two orders of magnitude at low compared to room temperature was measured. Moreover, we apply up to $\sim 400\,$mV compared to a maximum bias voltage of $\sim 50\,$mV in Ref. \cite{Arakawa2012}. Usually, $\alpha_H$ decreases with higher applied voltages by up to one order of magnitude\cite{Almeida2008,Gokce2006,Aliev2007} due to new conductance channels arising in this range of biases (and the likewise decreasing TMR ratio) \cite{Gokce2006,Aliev2007}. In our devices, this behaviour leads to a suppression of the phase noise increase for higher currents, as present in fig. \ref{fig_theo_prefactors:prefactors}. 
Finally, the parameter $\alpha_H$ in general depends on the magnetic configuration of the magnetic tunnel junction (in sensors mainly parallel P and antiparallel AP) because of the different nature of tunneling channels. 
The higher noise level in the AP state can (Ref. \cite{Scola2007}) be mostly explained by a higher contribution to noise from tunneling electrons from the localized bands. Those are more sensitive to charge traps (e.g. magnetic impurities) in the insulator. In the AP state, mainly localized bands contribute to the tunneling compared to mainly delocalized $s$-electrons in the P-state. Therefore, the difference in the Hooge parameter $\alpha_{H,AP}>\alpha_{H,P}$ is usually at least a factor of two \cite{Fermon2013} but more often even one or two orders of magnitude \cite{Herranz2011,Almeida2008,Aliev2007}.\\

As an important result, we state that the system's nonlinear nature strongly governs not only the thermal, but also the flicker noise behaviour. 
The influence on the statically observed noise proportionality $P_{dc}/A$, which is modelled by the Hooge-formalism, is found to be of minor significance at least until the larger current regime where the nonlinearity decreases and oscillation radius $s_0$ and oscillation output power $p_0$ do not significantly increase anymore. In this regime, the phase noise follows a Hooge-like increase, as we find in the evolution of the calculated parameters. Note however that this trend is found to be much less pronounced in the parameters extracted from the measurements (see fig. \ref{fig:prefactors_vs_current}) due to the evolution of $\alpha_H$ expected in this range of very large applied biases. 
In fact, the phenomenological parameter $\alpha_H$ describes the noise level and may have contributions from both magnetic and electronic mechanisms (as well as the microstructural quality of the thin films \cite{Eklund2014}) that might differ for amplitude and phase. A comparison with known values in similar TMR devices usually operating at very low voltage bias is not straightforward, because as we show here, all noise mechanisms might act in fundamentally different ways.
As we demonstrate, the evolution of the output power $p_0$ and the (correlated) active magnetic surface $A$ are essential for the phase noise.
For the amplitude noise, the essential governing parameter is the damping rate $f_p$ back to the limit cycle.\\

\section{Conclusion}

We present here a phenomenological theory based on the nonlinear auto-oscillator model in order to describe the noise properties of spin torque nano oscillators in the low offset frequency range.
We investigate how the $1/f$ flicker noise, which is the dominant noise source in the low offset frequency regime, is connected with the oscillator's nonlinear dynamics.

In addition to the new theoretical description, we also conduct a detailled experimental investigation performed on vortex based spin torque oscillators, and discuss the results based on the theoretical predictions. We measure a $1/f^1$ flicker amplitude noise and its conversion into $1/f^3$ phase noise. Indeed, the phase noise additionally exhibits a pure phase flicker noise contribution, which is found to be dominant. As a consequence, we hence demonstrate that the amplitude-phase-noise coupling is less important in the low offset frequency regime compared to higher frequency offsets.

In summary, we find that the flicker noise in spin torque oscillators, particularly in STVOs but extendable to other STNOs, is determined by the system's nonlinear parameters in the highly frequency tunable operation range. We also show that this conclusion becomes less valid in the regime of very large driving force on the vortex core (i.e. large applied current).
Moreover, we combine the dynamical equations with the well established Hooge formalism for TMR sensors and emphasize the importance of different parameters on the low offset frequency noise.

We believe that our work contributes to the fundamental understanding of the $1/f$ flicker noise in spin torque nano oscillators.
As this noise type especially limits the spin torque oscillator's long term stability, we furthermore believe that this work provides a basis to improve the noise characteristics of various applications aiming to rely on STNO devices.

\section{Acknowledgment}

S.W. acknowledges financial support from Labex FIRST-TF. The work is supported by the ANR SPINNET.

\bibliography{literatur_promo}

\end{document}